\documentclass[reprint,showpacs,preprintnumbers,amsmath,amssymb,prl]{revtex4-1}
\usepackage{amsmath}
\usepackage{amsfonts}
\usepackage{amsthm}
\usepackage{dsfont}
\usepackage{graphics}
\usepackage{graphicx}
\usepackage{subfigure}
\usepackage{amssymb}
\usepackage {mathrsfs}
\usepackage{braket}

\usepackage{color}
\usepackage{url}
\usepackage[abs]{overpic}
\usepackage[usenames,dvipsnames]{xcolor}

\begin{document}

\title{Action Principles for Quantum Automata and Lorentz Invariance of Discrete Time Quantum Walks}

\author{Fabrice Debbasch}
\email{fabrice.debbasch@gmail.com}
\affiliation{LERMA, UMR 8112, UPMC and Observatoire de Paris, 61 Avenue de l'Observatoire  75014 Paris, France}
\date{\today}
\begin{abstract}
A discrete action principle for general quantum automata is proposed. This action principle is particularized to Discrete Time Quantum Walks (DTQWs) and then extended into an energy and momentum preserving, manifestly covariant formulation. Space-time coordinates are introduced as new variables of the action and their equations of motion enforce energy and momentum conservation. This guarantees that the proposed action can be used to build future, DTQW-based self-consistent models of spinors interacting with gauge fields. A discrete stress-energy tensor for the DTQW is also obtained by functional differentiation of the action with respect to the gradients of the coordinates viewed as functions of the discrete grid points. The manifest covariance of the formulation highlights the special role played by the grid reference frame in the DTQW dynamics. The main discussion is complemented by three appendices.
\end{abstract}
\maketitle
\section{Introduction}

Quantum walks (DTQWs) are unitary quantum automata that can be viewed as formal generalizations of classical random walks. Following the seminal work of Feynman \cite{FeynHibbs65a} and Aharonov \cite{ADZ93a} they were considered in a systematic way by Meyer \cite{Meyer96a}. DTQWs have been realized experimentally with a wide range of physical objects and setups \cite{Schmitz09a, Zahring10a, Schreiber10a, Karski09a, Sansoni11a, Sanders03a, Perets08a}, and are studied in a large variety of contexts, ranging from quantum optics \cite{Perets08a}
to quantum algorithmics \cite{Amb07a, MNRS07a}, condensed matter physics \cite{Aslangul05a, Bose03a, Burg06a, Bose07a,DFB15a} and biophysics \cite{Collini10a, Engel07a}.

It has been shown recently \cite{CW13a,DBD13a,DBD14a,AFF15a,AF16a,Bru16a,AF16b,AP16a,AD17a} that several DTQWs can be considered as discrete models of Dirac fermions coupled to arbitrary Yang-Mills gauge fields (including electromagnetic fields) and to relativistic gravitational fields. And a DTQW coupled to a uniform electric field has already been realized experimentally \cite{MA13a}. It is thus tempting to think one could use DTQWs to build new self-consistent discrete models of Dirac fermions interacting with gauge fields, where DTQWs are not only acted upon by gauge fields, but also as sources to these fields. 
Ideally, such models would have to conserve energy and momentum of the whole system constituted of the DTQWs and the gauge field, and they also would have to be exactly both gauge invariant and covariant. If several DTQWs displaying exact discrete Yang-Mills gauge invariance have already been proposed, the only self-consistent interaction model \cite{AF16b}  which has been suggested for an electromagnetic field 
coupled to a DTQW does not conserve energy \cite{LQ16a} and Lorentz invariance of DTQWs is still a debated issue \cite{AFF14a,BDP16a,BDP17a}. 

The aim of this article is to kill these two birds with one stone in constructing manifestly covariant, energy and momentum preserving action principles for DTQWs which can be readily extended to include covariant kinetic terms for the gauge fields. 

We first prove that general unitary quantum automata admit a simple energy preserving action principle. We then focus on DTQWs with constant coefficients on $(1 + 1)$D space-time grid and use this general action principle to construct densities for the energy and momentum and for their fluxes in the proper frame of the $2$D space-time grid. 

This action principle, as it stands, presents two limitations. First, energy conservation derives from a particular algebraic property of unitary quantum automata which more general principles involving gauge fields will generically not exhibit, if only because gauge fields are not necessarily charged. Second, Lorentz invariance cannot be properly discussed on this action principle because it does not involve space-time coordinates, but only the discrete labels of grid points.

We therefore construct step by step a new action principle with space-time coordinates as extra variables. This action principle can be written in a manifestly covariant manner and be readily generalized to energy and momentum conserving action principles for DTQWs and gauge fields because energy and momentum conservation is now as consequence of the equations of motion for the space-time coordinates. As a consequence, functional differentiation with respect to the gradients of the space-time coordinates (conceived as functions of the grid labels) delivers the stress-energy tensor of the DTQW.

The manifest covariance of the formulation highlights that the grid reference frame plays a special role in the DTQW dynamics and this reference frame enters the equations of motion by its time-like velocity field and its space-like orthogonal.

These results are carefully discussed in the last section, and complemented with three Appendices. The first one is a tutorial on discrete variational principles. The second one shows how charge conservation can be recovered from the first action principle The third one revisits the Lorentz invariance of the flat space-time $(1 + 1)$D Dirac equation, with special emphasis on geometry and with extensions to curved space-times in mind. All material presented in the Appendices is new.

\section{\label{sec:basicAP}Basic action principle}

\subsection{General Quantum Automata}

Consider a discrete time quantum automaton defined in a certain Hilbert space $\mathcal H$ by the general abstract evolution equation $\Psi_{j + 1} = U_j \Psi_j$, where $\Psi_j$ represents the state of the automaton at time $j$ and $U_j$ is a possibly time-dependent unitary operator. The Hilbertian product in $\mathcal H$ will be denoted by brackets. 

For discrete time quantum walks (DTQWs) defined on a certain discrete spatial set $L$ (typically a lattice or a graph), $\Psi_j$ represents the collection $\left\{\psi^\sigma_{j p}, \sigma \in \Sigma, p \in L\right\}$ where $\sigma$ labels spin components and $p$ labels points in $L$. The action of the operator $U_j$ on $\Psi_j$ takes the typical form:
\begin{equation}
\left(U_j \Psi_j\right)^\sigma_p = \sum_{\sigma' \in L, q \in N_p} \left(U_{jp}^q \right)^\sigma_{\sigma'} \psi^{\sigma'}_{j q} 
\end{equation}
where the summation extends to all spin components and to points $q$ located in a certain `neighborhood' $N_p$ of point $p$. 

The Hilbertian product is then
\begin{equation}
< \Phi_j \mid \Psi_k > =  \sum_{\sigma \in L, p \in L} \left( \phi^\dagger_\sigma\right)_{j p} \psi^\sigma_{k p}.
\end{equation}

Consider now the action
\begin{equation}
S [\Psi, \Psi^\dagger] = \sum_j <\Psi_{j + 1} \mid \Psi_{j + 1} - U_j \Psi_j >
\label{eq:Mainaction}
\end{equation}
viewed as a functional of the independent variables $\Psi = \left\{\Psi_j, j \in \mathbb{Z} \right\}$ and  $\Psi^\dagger = \left\{\Psi^\dagger_j, j \in \mathbb{Z}\right\}$. 
Varying $\Psi$ and $\Psi^\dagger$ independently leads to
$\delta S = \Delta S + \Delta^\dagger S$ with
\begin{equation}
\Delta S
=  \sum_j  <\Psi_{j + 1} \mid \delta \Psi_{j + 1} - U_j \delta \Psi_j > 
\end{equation}
and 
\begin{equation}
\Delta^\dagger S = <\delta \Psi_{j + 1} \mid \Psi_{j + 1} - U_j \Psi_j > .
\end{equation}
Now,
\begin{eqnarray}
\Delta S
& =  & \sum_j  [<\Psi_{j + 1} \mid \delta \Psi_{j + 1}>  - <\Psi_{j+1} \mid U_j \delta \Psi_j > ]\nonumber \\
& = & \sum_j  <\Psi_{j} \mid \delta \Psi_{j}>  - \sum_j <U_j^\dagger \Psi_{j+1} \mid  \delta \Psi_j > \nonumber \\
& = & \sum_j  <\Psi_{j} - U_j^\dagger \Psi_{j+1}\mid \delta \Psi_{j}> \nonumber \\ 
& = &  \sum_j  <U_j^\dagger\left(U_j\Psi_{j} - \Psi_{j+1} \right)\mid \delta \Psi_{j}>
\end{eqnarray}
where the relation $U_j^\dagger U_j = 1$ expressing the fact that $U_j$ is unitary has been used to obtain the last equation.

Imposing that $S$ is extremal with respect to arbitrary variations of both $\Psi$ and $\Psi^\dagger$ therefore leads to the single equation $\Psi_{j + 1} =  U_j \Psi_j$. The functional $S$ can thus be used as action for the quantum automaton defined by $\left\{U_j, j \in \mathbb{Z} \right\}$. 
Note that the action (\ref{eq:Mainaction}) is clearly invariant under the global phase transformation $\Psi \rightarrow \exp(i \phi_0) \Psi$ with constant $\phi_0$ and thus conserves the charge 
$Q_j = < \Psi_j \mid \Psi_j >$.


\subsection{Conserved Hamiltonian}

Let us introduce the discrete derivative $\nu_j = (D \Psi)_j = \Psi_{j+1} - \Psi_j$ and rewrite the action $S$ as
\begin{equation}
{\tilde S} [\Psi, \Psi^\dagger, \nu] = \sum_j <\Psi_{j + 1} \mid \nu_{j} - (U_j - 1) \Psi_j >,
\end{equation}
where $\Psi^\dagger_{j+1}$ has not been expressed in terms of $\nu^\dagger_j$ to ensure that all discrete equations follow as closely as possible their continuous counterpart valid for the Dirac equation.

The momentum of $\Psi_j$ is $\pi_j = \Psi_{j+1}^\dagger$
and the momentum of $\Psi^\dagger_j$ vanishes identically. The Legendre transform of ${\tilde S}_H$ reads:
\begin{equation}
{\tilde S}_H [\Psi, \pi] = \sum_j <\pi^\dagger_j \mid (U_j - 1) \Psi_j>.
\end{equation}
This function delivers the correct equation of motion in the form 
\begin{equation}
(D \Psi)_j = \Psi_{j+1} - \Psi_j = +\left( \frac{ \partial {\tilde S}}{\partial \pi}\right)_{\mid j} = (U_j - 1) \Psi_j.
\end{equation}
and
\begin{equation}
(D \pi)_{j-1} = \pi_{j} - \pi_{j-1} = - \left(\frac{ \partial {\tilde S}}{\partial \Psi}\right)_{\mid j} = - \pi_j (U_j - 1)
\end{equation}
This second equation can be rewritten $- \pi_{j-1} = - \pi_j U_j$ or $\pi_{j-1}^\dagger =  U_j^\dagger \pi_j^\dagger$ or, using the unitarity of $U_j$:
\begin{equation}
U_j\pi_{j-1}^\dagger = \pi_j^\dagger,
\end{equation}
which admits $\pi_j^\dagger = \Psi_{j+1}$ as solution. 

Contrary to the general discrete case (see Appendix), the Hamiltonian $H_j (\Psi_j, \pi_j) = <\pi^\dagger_j \mid (U_j - 1) \Psi_j>$ {\sl is} then conserved if $U_j = U$ does not depend explicitly on $j$. Indeed, one can then write
\begin{eqnarray}
H_{j+1} & = & <\pi^\dagger_{j+1} \mid (U - 1) \Psi_{j + 1}> \nonumber \\
& = & <U\pi^\dagger_{j} \mid (U - 1) U \Psi_{j}> \nonumber \\
& = & <\pi^\dagger_{j} \mid U^\dagger (U - 1) U \Psi_{j}> \nonumber \\
& = & <\pi^\dagger_{j} \mid (U - 1) \Psi_{j}> \nonumber \\
& = & H_j
\end{eqnarray}
where the unitarity of $U$ has been used in passing from the third to the fourth equation.

\subsection{\label{ssec:Energy}Energy Conservation for DTQWs on the line}

Let us now focus, for simplicity sakes, on the so-called one-step $(1 + 1)$D DTQWs with two component wave functions $\Psi = (\psi^-, \psi^+)^\dagger$, for which $U_j = V_j T$ where $T$ is the $j$-independent spatial translation operator $T$ defined by
\begin{eqnarray}
(T\psi)^{-}_{p} & = &  \psi^{-}_{p+1} \nonumber \\
(T \psi)^{+}_{p} & = &  \psi^{+}_{p-1} 
\end{eqnarray}
and $V_j$ is defined 
\begin{equation}
(V_j \Psi)_p = W_{j, p} \Psi_p
\end{equation}
with $W_{j,p}$ an arbitrary $j$- and $p$-dependent operator in $U(2)$.

It is useful, for DTQWs on the line, to consider the following alternate form of the action $\tilde S$:
\begin{equation}
{\tilde S} [\Psi, \Psi^\dagger, \nu] = \sum_j <\Psi_{j} \mid U^\dagger \nu_{j} - (1- U^\dagger) \Psi_j >,
\end{equation}
which can be obtained form the original form ( ) by taking into account that $\Psi_{j + 1} = U \Psi_j$. This alternate form delivers the Hamiltonian under the form
$ H_j = <\Psi_{j} \mid (1 - U^\dagger_j) \Psi_j>$. A straightforward computation shows that the corresponding energy density 
\begin{equation}
{\mathcal H}_{j, p} = \Psi^\dagger_{j, p} \left( (1 - U^\dagger) \Psi_j\right)_p
\end{equation}
obeys the conservation equation
\begin{equation}
\left( \nabla_j {\mathcal H} \right)_{j, p} + \left( \nabla_p {\mathcal J}_{\mathcal H} \right)_{j, p} = 0
\end{equation}
with 
\begin{equation}
({\mathcal J}_{\mathcal H})_{j, p} = - \Psi^\dagger_{j, p} \sigma_3 \left( (1 - U^\dagger)  \Psi_j\right)_p
\end{equation}
where $\sigma_3$ is the third Pauli matrix {\sl{i.e.}} $\sigma_3 = \mbox{diag} (1, -1)$ and
\begin{equation}
\left( \nabla_j {\mathcal H} \right)_{j, p} = {\mathcal H}_{j+1, p} - \frac{{\mathcal H}_{j, p+1} + {\mathcal H}_{j, p-1}}{2}, 
\end{equation}
\begin{equation}
\left( \nabla_p {\mathcal J}_{\mathcal H} \right)_{j, p}  = \frac{({\mathcal J}_{\mathcal H})_{j, p+1} 
- ({\mathcal J}_{\mathcal H})_{j, p-1}
}{2}.
\end{equation}

Note that ${\mathcal H}_{j, p} = Q_{j, p} + h_{j, p}$, where $Q_{j, p} = \Psi^\dagger_{j, p} \Psi_{j, p}$ is the local charge density and $h_{j, p} = - \Psi^\dagger_{j, p} \left(U^\dagger \Psi_j\right)_p$. The current $({\mathcal J}_{\mathcal H})_{j, p}$ accordingly splits into two contributions, the charge current $ ({\mathcal J}_{Q})_{j, p} = - \Psi^\dagger_{j, p} \sigma_3  \Psi_{j, p}$ (\cite{AF16b}) and the current $({\mathcal J}_h)_{j, p} =  \Psi^\dagger_{j, p} \sigma_3 \left(U^\dagger \Psi_j\right)_p$.


We now define the momentum $\mathcal P$ of the DTQW by
\begin{equation}
{\mathcal P}_{j, p} =  \frac{1}{2}\ \Psi^\dagger_{j, p} \sigma_3 \left( (T - T^\dagger) \Psi_j\right)_p.
\end{equation}
This quantity is conserved if all coefficients of the DTQW are constants, as can be easily seen by writing the DTQW evolution equations in Fourier space, and also coincides, in the continuous limit, with the wave number of the walk. The associated momentum current is
\begin{equation}
({\mathcal J}_{\mathcal P})_{j, p} = -  \frac{1}{2}\ \Psi^\dagger_{j, p}  \left( (T - T^\dagger) \Psi_j\right)_p.
\end{equation}


In working with the alternate form of the action, it is useful to change 
variables {\sl i.e.} to replace the independent variables $\Psi$ and $\Psi^\dagger$ by the variables
$\Phi = U^\dagger \Psi$ and $\Psi^\dagger = (U \Phi)^\dagger$. The action then reads:
\begin{equation}
{\tilde S} [\Phi, \Psi^\dagger, \nu^\Phi] = \sum_j <\Psi_{j} \mid \nu^\Phi_{j} - (U- 1) \Phi_j >,
\label{eq:Mainaction1}
\end{equation}
where $\nu^\Phi_j = \Phi_{j + 1} - \Phi_j$ is the new velocity. 


%


\section{\label{eq:Extended}Extended Action Principle for DTQWs on the line}

The dynamics derived from the above action principle conserves the Hamiltonian $H$ of quantum automata and also the impulse $P$ of DTQWs. However, as it stands, it cannot be generalized into action principles ensuring energy and momentum conservation for discrete gauge theories based on DTQWs \cite{AF16b,ADBD16a}. As explained in Appendix 1, the solution to the problem consists in extending the basic action principle by introducing continuous space-time coordinates defined on the lattice. We will now do so for DTQWs on the line, restricting the discussion to walks with constant coefficients {\sl i.e.} constant $W$ operator. This will deliver an extended action principle from which a conserved stress-energy `tensor' can be computed. It turns out, as a bonus, that the extended action principle is invariant under a generalization of the Lorentz transform. The case of DTQWs with non constant coefficients will be dealt with in a subsequent publication.

\subsection{\label{ssec:Intrinsic}Intrinsic geometry on the grid}

We first make explicit the fact that the operator $T$ codes for spatial discrete derivatives by writing
\begin{equation}
T = \sigma_3 \nabla_p + C
\label{eq:Tsplit}
\end{equation}
where
\begin{equation}
\left(\nabla_p \Phi \right)_{j, p} = \frac{1}{2}\  \left( \Phi_{j, p+1} - \Phi_{j, p-1} \right) 
\end{equation}
and
\begin{equation}
\left(C \Phi \right)_{j, p} = \frac{1}{2}\  \left( \Phi_{j, p+1} + \Phi_{j, p-1} \right).
\end{equation}
In Fourier space for a wave vector $k$, the operator $\nabla_p$ delivers a $i \sin k$ factor, which becomes $i k$ {\sl i.e.} a derivation in the continuous limit, and the operator $C$ delivers a $\cos k$ factor, which becomes unity in the same limit.

We also introduce the discrete time derivative operator $D_j$ defined by 
\begin{equation}
\left(\nabla_j \Phi \right)_{j, p} = \left( \Phi_{j + 1, p} - \Phi_{j,  p} \right) = (\nu_j^\Phi)_{j, p}
\end{equation}
and use the definition $U = WT$ to write the action ${\tilde S}$ under the form
\begin{eqnarray}
{\tilde S} [\Phi, \Psi^\dagger, \nabla_j\Phi, \nabla_p \Phi] & = & \sum_j <\Psi_{j} \mid (\nabla_j \Phi)_{j} - W \sigma_3 (\nabla_p \Phi)_j  \nonumber \\
& - & (WC - 1) \Phi_j >_p,
\label{eq:Mainaction2}
\end{eqnarray}
where the notation $< A \mid B>_p = \sum_p A^*_p B_p$ has been used.  

The conserved Hamiltonian $\mathcal H$ reads
\begin{eqnarray}
{\mathcal H}_j & = & < \Psi_j \mid (U - 1) \Phi_j >_p \nonumber \\
& = & < \Psi_j \mid (W \sigma_3 \nabla_p + WC - 1) \Phi_j>_p
\end{eqnarray}
and the conserved momentum $\mathcal P$ reads
\begin{eqnarray}
{\mathcal P}_j & = & \frac{1}{2} \ <\Psi_j \mid \sigma_3 (T - T^\dagger) U \Phi_j>_p \nonumber \\
& = & <\Psi_j \mid \nabla_p U \Phi_j>_p \nonumber \\
& = & <\Psi_j \mid U \nabla_p \Phi_j>_p.
\end{eqnarray}

\subsection{\label{ssec:STcoord}Space-time coordinates}

The extended action $S_e$ involves 
two
additional variables. 
These
are a continuous time variable $t_{j, p} = X^0_{j, p}$ and a continuous space variable $x_{j, p} = X^1_{j, p}$. If one follows the procedure outlined in Appendix 1, a key step in obtaining the action $S_e$ from $S$ is to replace $\nabla_j$ and $\nabla_p$ by discrete versions $\nabla_t = \nabla_0$ and $\nabla_x = \nabla_1$ of the time- and space-derivatives. These are constructed as follows. First suppose that $j$ and $p$ were actually continuous coordinates. One could then write
\begin{eqnarray}
\frac{\partial}{\partial t} & = & \frac{\partial j}{\partial t} \frac{\partial}{\partial j} + \frac{\partial p}{\partial t} \frac{\partial}{\partial p} \nonumber \\
\frac{\partial}{\partial x} & = & \frac{\partial j}{\partial x} \frac{\partial}{\partial j} + \frac{\partial p}{\partial x} \frac{\partial}{\partial p}.
\end{eqnarray}
The coefficients in front of $\frac{\partial}{\partial j}$ and $\frac{\partial}{\partial p}$ are those of the Jacobian $\partial (j, p)/\partial(t, x)$. This Jacobian is the inverse of $\partial (t, x)/\partial(j, p)$, form which one 
deduces
\begin{eqnarray}
\frac{\partial j}{\partial t} & = &  + \frac{1}{\delta}  \frac{\partial x}{\partial p} \nonumber \\
\frac{\partial j}{\partial x} & = &  - \frac{1}{\delta} \frac{\partial t}{\partial p} \nonumber \\  
\frac{\partial p}{\partial t} & = &  - \frac{1}{\delta}  \frac{\partial x}{\partial j}  \nonumber \\
\frac{\partial p}{\partial x} & = &  +  \frac{1}{\delta} \frac{\partial t}{\partial j}
\end{eqnarray}
where 
\begin{equation}
\delta = \frac{\partial t}{\partial j} \frac{\partial x}{\partial p} - \frac{\partial t}{\partial p} \frac{\partial x}{\partial j}.
\end{equation}

This suggests that the building blocks from which an extended, coordinate dependent DTQW action can be constructed are
\begin{eqnarray}
C^j_0 & = & + \frac{1}{\Delta (\nabla X)}  \nabla_p X^1 \nonumber \\
C^p_0 & = & - \frac{1}{\Delta (\nabla X)}  \nabla_j X^1 \nonumber \\
C^j_1 & = & - \frac{1}{\Delta (\nabla X)}  \nabla_p X^0 \nonumber \\
C^p_1 & = & + \frac{1}{\Delta (\nabla X)}  \nabla_j X^0
\label{eq:defC}
\end{eqnarray}
with
\begin{equation}
\Delta  (\nabla X) = (\nabla_j X^0) (\nabla_p X^1) - (\nabla_p X^0) (\nabla_j X^1)
\end{equation}
where
\begin{eqnarray}
\left( \nabla_j X^\mu \right)_{j, p} &  = & \frac{X^\mu_{j, p-1} + X^\mu_{j, p + 1}}{2}  - X^\mu_{j-1, p} \nonumber \\
\left( \nabla_p X^\mu \right)_{j, p} & =  & \frac{X^\mu_{j, p+1} - X^\mu_{j, p-1}}{2}
\end{eqnarray} 

One could hope that replacing $\nabla_j$ ({\sl resp.} $\nabla_p$) by $\nabla_0 = C^j_0 \nabla_j + C^p_0 \nabla_p$ ({\sl resp.} $\nabla_1 = C^j_1 \nabla_j + C^p_1 \nabla_p$) in the original action 
(\ref{eq:Mainaction2}) delivers a correct, coordinate-dependent extended action, but this simple procedure turns out to be a bit too naive. The right extended action is presented in the next section.

\subsection{Action with coordinates and conserved quantities}

Consider the following action:
\begin{equation}
{\Sigma} [\Phi, \Psi^\dagger, \nabla \Phi, \nabla X]
= \sum_j \left(K_{j} + M_j \right)
\label{eq:Coordaction1}
\end{equation}
where the `mass' term is
\begin{equation}
M_j = M^1_j + M^2_j + M^3_j
\end{equation}
with
\begin{equation}
M^1_j =  < \Psi_j \mid \left(1 - WC\right) \frac{1 + T}{2} \Delta \left((\nabla X)_j\right) \Phi_j>_p,
\end{equation}
\begin{equation}
M^2_{j} = < \Psi_j \mid \left( 1 - WC \right) \frac{1 - T}{2} \left( (\nabla_j X^0)_j - (\nabla_p X^1)_j \right)
\Phi_j >_p,
\end{equation}
\begin{equation}
M^3_j =  < \Psi_j \mid \left( 1 - WC \right) \frac{1 - T}{2} \Phi_j >_p,
\end{equation}
and the more complicated `kinetic' term is
\begin{equation}
K_{j} = K^j_{j} + K^p_{j} + K^{\mbox{\tiny supp}}_{j}
\end{equation}
with
\begin{equation}
K^j_ {j} = - < \Psi_j \mid W \sigma_3 (\nabla_j \chi)_j  (\nabla_p \Phi)_j >_p,
\end{equation}
\begin{equation}
K^p_{j} = + < \Psi_j \mid \sigma_3  (\nabla_p \chi)_j  (\nabla_j \Phi)_j >_p,
\end{equation}
\begin{equation}
K^{\mbox{\tiny supp}}_{j} = + < \Psi_j \mid (1 - T)  (\nabla_j \Phi)_j >_p
\end{equation}
where 
\begin{equation}
\chi = X^0 \mathds{1} + X^1 \sigma_3 T.
\end{equation}
The unitary operators $W$ and $T$ are those entering the definition of the DTQW given above.

The first point to note is that this action coincides 
with the action (\ref{eq:Mainaction2}) for $(\Phi, \Psi)$ when $X^0_{j, p} = j$ and $X^1_{j, p} = p$. Thus, with this choice of coordinates, the action $\Sigma$ gives back the usual 
DTQW dynamics. Note that $M^2_j$ vanishes identically when $X^0_{j, p} = j$ and $X^1_{j, p} = p$. Let us now investigate the equations of motion for $X^0$ and $X^1$ derived from $\Sigma$.

The functional derivatives of $\Sigma$ with respect to the gradients of the 
coordinates read:
\begin{eqnarray}
\frac{\delta \Sigma}{\delta \left( (\nabla_j X^0)_{j, p}  \right)} & = & \Psi^\dagger_{j, p} \Big( - W \sigma_3 (\nabla_p \Phi)_{j, p} \Big. \nonumber \\
& + & \left. (1 - WC) \frac{1 - T}{2}\  \Phi_{j, p}\right. \nonumber\\
& + & \Big. (1 - WC) \frac{1 + T}{2}\  (\nabla_p X^1)_{j, p} \Phi_{j, p}\Big), 
\end{eqnarray}
\begin{eqnarray}
\frac{\delta \Sigma}{\delta \left( (\nabla_p X^0)_{j, p}  \right)} & = & \Psi^\dagger_{j, p} \Big( \sigma_3 (\nabla_j \Phi)_{j, p} \Big. \nonumber\\
& - & \Big.  (1 - WC) \frac{1 + T}{2}\ (\nabla_j X^1)_{j, p}  \Phi_{j, p}\Big), 
\end{eqnarray}
\begin{eqnarray}
\frac{\delta \Sigma}{\delta \left( (\nabla_j X^1)_{j, p}  \right)} & = & \Psi^\dagger_{j, p} \Big( - U (\nabla_p \Phi)_{j, p} \Big. \nonumber\\
& - & \Big.  (1 - WC) \frac{1 + T}{2}\ (\nabla_p X^0)_{j, p}  \Phi_{j, p}\Big), 
\end{eqnarray}
and
\begin{eqnarray}
\frac{\delta \Sigma}{\delta \left( (\nabla_p X^1)_{j, p}  \right)} & = & \Psi^\dagger_{j, p} \Big( T (\nabla_j \Phi)_{j, p} \Big. \nonumber\\
& - & \left.  (1 - WC)  \frac{1 - T}{2}\  \Phi_{j, p} \right. \nonumber \\
& + & \Big. (1 - WC)  \frac{1 + T}{2}\  (\nabla_j X^0)_{j, p} \Phi_{j, p}\Big).
\end{eqnarray}

Evaluating these quantities on shell {\sl i.e.} for $X^0_{j, p} = j$, $X^1_{j, p} = p$ and $(\nabla_j \Phi)_{j, p} = (U-1) \Phi_{j, p}$ leads to 
\begin{equation}
\frac{\delta \Sigma}{\delta \left( (\nabla_j X^0)_{j, p}  \right)} \mathrel{\widehat{=}}
- {\mathcal H}_{j, p},
\end{equation}
\begin{equation}
\frac{\delta \Sigma}{\delta \left( (\nabla_p X^0)_{j, p}  \right)}  \mathrel{\widehat{=}}
- ({\mathcal J_{\mathcal H}})_{j, p},
\end{equation}
\begin{equation}
\frac{\delta \Sigma}{\delta \left( (\nabla_j X^1)_{j, p}  \right)} \mathrel{\widehat{=}}
-  {\mathcal P}_{j, p}
\end{equation}
\begin{equation}
\frac{\delta \Sigma}{\delta \left( (\nabla_p X^1)_{j, p}  \right)} \mathrel{\widehat{=}}
-  ({\mathcal J}_{\mathcal P})_{j, p}
\end{equation}
where 
$\mathrel{\widehat{=}}$ has been used to designate on shell equalities.

This shows that the equations of motion for $X^0$ and $X^1$, when taken on shell, come down to the energy and momentum conservation laws for the DTQW. Thus the DTQW dynamics $(\nabla_j \Phi)_{j, p} = (U-1) \Phi_{j, p}$, together with the choice of coordinates $X^0_{j, p} = j$, $X^1_{j, p} = p$, solves all equations of motion derived from the extended action $\Sigma$.

\subsection{Lorentz invariance I: main kinetic term}

We first stress that the appearance of $X^0$ and $X^1$ in the action $\Sigma$ traces the fact that $\Sigma$ is an off-shell action, not that $\Sigma$ is valid in an arbitrary Lorentz frame. Indeed, because the equations of motion derived from $\Sigma$ are solved by $X^0 = j$ and $X^1 = p$, $\Sigma$ is actually valid in the rest frame of the grid only. Note that the conserved quantities derived from $\Sigma$ are also the energy and momentum in the rest frame of the grid. Our main task is thus to define Lorentz transformations for DTQWs in such a way that (i) these Lorentz transformations coincide with usual continuous Lorentz transformations in the continuous limit (ii) it is possible to find a covariant off-shell action which coincides with $\Sigma$ in the rest frame of the grid.

 The first step consists in identifying in $\Sigma$ a main kinetic term which ressembles the kinetic from the flat space-time Dirac equation. We will use this term to define Lorentz transformations on DTQWs and then introduce a covariant form for this main kinetic term.   
 
The second step, which will be presented in the next section, consists in dealing with the mass terms and with the extra kinetic terms which vanish in the continuous limit and thus do not contribute to the main kinetic term studied in this section.

The operator $W \sigma_3$ is in $U(2)$. It therefore admits two eigenvalues of unit modulus, which we call $\exp(i \alpha_L)$ and $\exp(i \alpha_R)$. We now switch to a new spin basis $(b_f) = (b_L, b_R)$ made of normalized eigenvectors of $W \sigma_3$. In this new basis, the operator $\sigma_3$ is not represented by the third Pauli matrix. We therefore introduce the new operator ${\bar \sigma}_3$ whose representation in 
$(b_f)$ coincides with the third Pauli matrix and write
$\sigma_3 T = \left( {\bar \sigma}_3 + \left(\sigma_3 - {\bar \sigma}_3 \right) \right) \left( 1 + \left( T - 1 \right) \right)$, which leads to 
$\chi =  {\bar \chi} + \Delta \chi$ with  
\begin{eqnarray}
{\bar \chi} & =  & X^0 + X^1 {\bar \sigma}_3 \nonumber \\
\Delta \chi & = & X^1 \left[  \left(\sigma_3 - {\bar \sigma}_3 \right)  + {\bar \sigma}_3 \left( T - 1 \right) 
\right. \nonumber \\
& & \left.  
+ \left(\sigma_3 - {\bar \sigma}_3 \right)  \left( T - 1 \right) \right]
\end{eqnarray}
Similarly, $\xi = \sigma_3 \chi = X^0 \sigma_3 + X^1 T = {\bar \xi} + \Delta \xi$ with
\begin{eqnarray}
{\bar \xi} & =  & X^0 {\bar \sigma}_3 + X^1 \nonumber \\
\Delta \xi & = & X^0 \left(\sigma_3 - {\bar \sigma}_3 \right)  + X^1 \left( T - 1 \right) 
\end{eqnarray}

Inserting the above expressions into the kinetic terms of the action $\Sigma$ delivers expressions of the form
$K^{j/p}_j = {\bar K}^{j/p}_j + {\Delta K}^{j/p}_j$ with
\begin{equation}
{\bar K}^j_ {j} = - < \Psi_j \mid W \sigma_3 (\nabla_j {\bar \chi})_j  (\nabla_p \Phi)_j >_p,
\end{equation}
\begin{equation}
{\bar K}^p_{j} = + < \Psi_j \mid (\nabla_p {\bar \xi})_j  (\nabla_j \Phi)_j >_p.
\end{equation}
The sum ${\bar K}_j = {\bar K}^{j}_j + {\bar K}^{p}_j$ can be written as ${\bar K}_j = \sum_p {\bar K}_{j, p}$ with 
\begin{eqnarray}
{\bar K}_{j, p} = \eta_{fg} (\Psi^f)^*_{j, p} \left[ \left( e^q_a \gamma^0 \gamma^a D_q \right)^g_h \Phi^h  \right]_{j, p} \Delta ((\nabla X)_{j, p})
\end{eqnarray}
with $D_j = \nabla_j$, $D_p = W \sigma_3 {\bar \sigma}_3 \nabla_p$, $e^q_a = C^q_a$ (see (\ref{eq:defC})),
\begin{eqnarray}
(\gamma^0)^f_g & = & 
\begin{pmatrix} 
0 & 1 \\
1 & 0 
\end{pmatrix}
\nonumber \\
(\gamma^1)^f_g & = &
\begin{pmatrix} 
0 & 1 \\
-1 & 0 
\end{pmatrix},
\end{eqnarray}
and summation is implied over repeated indices $\{f, g, h\} \in \{L, R\}^3$, $q \in \{j, p\}^2$ and $a \in \{0, 1\}$.
The coefficients $(e^q_a = C^q_a)$ can be interpreted as the $(j, p)$ `components' of an $X$-dependent $2$-bein
$(e_0, e_1)$ with $\mu$-components $e_0^0 = 1  = e_1^1$, $e_0^1 = e_1^0 = 0$ (see Section `Space-time 
coordinates' above as well as Appendix 3). Introducing $D_\mu = C_\mu^j D_j + C_\mu^p D_p$ helps bring
${\bar K}_{j, p}$ under the form 
\begin{equation}
{\bar K}_{j, p} = \eta_{fg} (\Psi^f)^*_{j, p} \left[ \left( e^\mu_a \gamma^0 \gamma^a D_\mu \right)^g_h \Phi^h  \right]_{j, p} \Delta ((\nabla X)_{j, p}),
\end{equation}
which is similar to the usual form of the kinetic term in the Dirac equation. We thus define the DTQW Lorentz transformation exactly as the continuous one and
we proceed
in the same exact way. In particular, the basis vector $b _L$ ({\sl resp.} $b_R$) 
gets multiplied by 
a positive real number $\lambda$ ({\sl resp.} $\lambda^{-1}$) and this change of basis is compensated by a change of coordinates through a change of spin metric and of $2$-bein. The volume `measure' $\Delta (\nabla X) \Delta j \Delta p = \Delta (\nabla X)$ is invariant under Lorentz transformation. Note that the Lorentz transformation of ${\bar K}_{j}$ is computationally simple because we are working in the basis which makes diagonal $W \sigma_3$, ${\bar \sigma}_3$, and thus also their product  $W \sigma_3 {\bar \sigma}_3$. Choosing the basis $(b_f)$ orthonormal ensures that the spin-metric is diagonal too.  

\subsection{Lorentz invariance II: extra kinetic terms and mass terms}

As a prolegomenon, let us remark that, as any quantity appearing in a covariant theory, an arbitrary spin operator can be fully defined 
by its value in a particular spin-basis together with its transformation law. This way of defining spin-operators will be used extensively in what follows where most spin operators will be defined by their values in the spin basis corresponding to the grid reference frame combined with the SOLT law introduced in Appendix 3 to discuss Lorentz invariance for the Dirac equation. 

There are three extra kinetic terms {\sl i.e.} $\Delta K^j$, $\Delta K^p$ and $K^{\mbox{supp}}$, and two mass terms {\sl i.e.} $M^1$ and $M^2$.
The mass terms are somewhat simpler and we start with them.

Since $\Delta ((\nabla X)_{j, p})$ is Lorentz invariant, the first mass-term $M^1_j = \sum_p M^1_{j, p}$ can be written in the covariant form 
\begin{equation}
M^1_{j, p} = \eta_{fg} (\Psi^f)^*_{j, p} \left( (m^1)^g_h \Phi^h \right)_{j, p} \Delta ((\nabla X)_{j, p})
\end{equation}
where the operator $m^1 = (1 - WC) \frac{1 + T}{2}$ in the spin basis of the grid frame and obeys the SOLT law. 

The second mass term is conceptually more complicated. To write in a manifestly covariant manner, let us introduce 
the `vector field' ({\sl{resp.}} `form field') $U^\mu$ ({\sl{resp.}} $U_\mu$) is defined by its components $U^0 = 1$, $U^1 = 0$ ({\sl{resp.}} $U_0 = 1$, $U_1 = 0$) in the grid frame, together with the usual Lorentz transformation law for vector fields ({\sl{resp.}} form-fields). In physical terms, $U$ represents the velocity field of the grid in $2D$ space-time. It also coincides with the first vector of the $2$-bein associated to the grid coordinates (see the definition 
of the coordinate-dependent $2$-beins in the previous section). Note that the grid being independent of the choice of coordinates, $\nabla_j$ behaves like a scalar. In usual continuous geometrical terms, $\nabla_j = U^\nu \nabla_\nu$. This scalar depends on $U$ {\sl i.e.} on the choice of grid, not on the choice of coordinates. 
We also introduce the `vector field' $V^\mu$ defined by its components $V^0 = 0$ and $V^1 = 1$ in the grid frame, together with its associated form $V_\mu$. This vector field coincides with the second vector of the $2$-bein associated to the grid coordinates and will be used immediately below to express one of the extra kinetic terms in a manifestly covariant manner. And the operator $\nabla_p$ is a scalar, discrete equivalent to $V^\nu \nabla_\nu$. Note that contracting a tensor with $V$ essentially amounts to projecting the tensor on the orthogonal to $U$:
\begin{eqnarray}
\Pi_{\mu \nu} A^\nu & = & (\eta_{\mu \nu} - U_\mu U_\nu) A^\nu \nonumber \\
& = & (\eta_{\mu \nu} - U_\mu U_\nu) \left( (A.U) U^\nu + (A.V) V^\nu \right) \nonumber \\
& = & (A.V) V_\mu,
\end{eqnarray}
where $\eta_{\mu \nu}$ is the Minkovski metric in orthonormal coordinates, $\Pi$ is the projector unto the orthornomal to $U$ and $(A.B) = 
\eta_{\alpha \beta} A^\alpha B^\beta$.

The `vector fields' $U$ and $V$ help write $M^2_j = \sum_p M^2_{j, p}$ with
\begin{equation}
M^2_{j, p} = \eta_{fg} (\Psi^f)^*_{j, p} \left( (m^2)^g_h  \Phi^h \right)_{j, p}  (U_\mu \nabla_j  + V_\mu \nabla_p)X^\mu
\end{equation}
where the spin-operator $m^2$ obeys the SOLT law and coincides with $ (1 - WC) \frac{1 - T}{2}$ in the spin basis of the grid frame.


Similarly, $M^3_j = \sum_p M^3_{j, p}$ with
\begin{equation}
M^3_{j, p} = \eta_{fg} (\Psi^f)^*_{j, p} \left( (m^2)^g_h  \Phi^h \right)_{j, p}.
\end{equation}


The three extra kinetic terms can be dealt with in the same manner. Their Lorentz invariant expressions are
$K^{\mbox{\tiny supp}}_j = \sum_p K^{\mbox{\tiny supp}}_{j, p}$, $\Delta K^{j}_j = \sum_{p} \Delta K^{j}_{j, p}$, $\Delta K^{p}_j = \sum_{p'} \Delta K^{p}_{j, p'}$, 
with 
\begin{equation}
K^{\mbox{\tiny supp}}_{j, p} = \eta_{fg} (\Psi^f)^*_{j, p} \left( (k^{\mbox{\tiny supp}})^g_h  \nabla_j \Phi^h \right)_{j, p} 
\end{equation}
\begin{eqnarray}
\Delta K^{j}_{j, p} & = & + \eta_{fg} (\Psi^f)^*_{j, p} \left( (\Delta k^{j})^g_h  \nabla_p \Phi^h \right)_{j, p} \nonumber \\
& & \times (V_\mu \nabla_j  X^\mu)
\end{eqnarray}
\begin{eqnarray}
\Delta K^{p}_{j, p} & = & \eta_{fg} (\Psi^f)^*_{j, p} \left( (\Delta k_U^{p})^g_h  \nabla_j \Phi^h \right)_{j, p} 
\nonumber \\
& & \times (U_\mu \nabla_p  X^\mu)
\nonumber \\
& - &  \eta_{fg} (\Psi^f)^*_{j, p} \left( (\Delta k_V^{p})^g_h  \nabla_j \Phi^h \right)_{j, p} \nonumber \\
& & \times (V_\mu  \nabla_p  X^\mu)
\end{eqnarray}
where the operators $k^{\mbox{\tiny supp}}$, $\Delta k^{j}$, $\Delta k_U^{p}$, $\Delta k_V^{p}$ all obey the SOLT law and coincide respectively with 
$1 - T$, $W \sigma_3 \left[  \left(\sigma_3 - {\bar \sigma}_3 \right)  + {\bar \sigma}_3 \left( T - 1 \right) + \left(\sigma_3 - {\bar \sigma}_3 \right)  \left( T - 1 \right) \right]$, $\sigma_3 - {\bar \sigma}_3$ and $T - 1$ in the spin basis of the grid frame. We have also used the standard notations $U_\mu = \eta_{\mu \nu} U^\mu$ and $V_\mu = \eta_{\mu \nu} V^\mu$ with $\eta_{\mu \nu} = \mbox{diag} (1, -1)$, in accordance with the $2$-bein components introduced in 
the previous section. 

\subsection{Stress-energy in an arbitrary Lorentz frame}

Let $\Sigma_L$ be the Lorentz invariant action obtained by summing up all previous contributions. By construction, the on-shell values of the functional derivatives of $\Sigma_L$ with respect to the derivatives of the coordinates coincide, in the grid reference frame, with the (opposite of) the Hamiltonian $\mathcal H$, the momentum $\mathcal P$, and their currents ${\mathcal J}_{\mathcal H}$ and ${\mathcal J}_{\mathcal P}$. These four quantities represent the stress-energy of the DTQW in the grid frame. Because we want to investigate what happens to the stress-energy in other Lorentz frames, we view these quantities as the grid frame components of a single object, the stress-energy `tensor' $\mathcal T$ of the walk. Since the on-shell values of the coordinates in the grid reference frame are $X^0 = j$ and $X^1 = p$, we denote the components of the stress energy `tensor' in the grid frame by ${\mathcal T}_j^j = \mathcal H$, ${\mathcal T}_j^p = {\mathcal J}_{\mathcal H}$,
 ${\mathcal T}_p^j = \mathcal P$ and ${\mathcal T}_p^p = {\mathcal J}_{\mathcal P}$. As shown above, these components obey the two conservation equation $\nabla_j {\mathcal T}_j^j  + \nabla_p {\mathcal T}_j^p = 0$ and
$\nabla_j {\mathcal T}_p^j  + \nabla_p {\mathcal T}_p^p = 0$.

Suppose now we change Lorentz frame to coordinates which we denote by $X^0$ and $X^1$ for simplicity sake. It is straightforward to show that the on-shell values of the derivatives of $\Sigma_L$ with respect to the coordinate derivatives coincide on-shell with the Lorentz transforms of the components ${\mathcal T}_{j/p}^{j/p}$ with respect to the lower index (remember that the $C$'s are the coefficients to be applied on forms when one changes form coordinates ${j, p}$ to coordinates $(X^0, X^1)$, see above):
\begin{eqnarray}
{\mathcal T}_0^j & = & - \frac{\delta \Sigma}{\delta \left( \nabla_j X^0\right)} \mathrel{\widehat{=}} C_0^j {\mathcal T}_j^j + 
C_0^p {\mathcal T}_p^j  \nonumber \\
{\mathcal T}_0^p  & = & - \frac{\delta \Sigma}{\delta \left( \nabla_p X^0\right)} \mathrel{\widehat{=}} C_0^j {\mathcal T}_j^p + 
C_0^p {\mathcal T}_p^p  \nonumber \\
{\mathcal T}_1^j  & = & - \frac{\delta \Sigma}{\delta \left( \nabla_j X^1\right)} \mathrel{\widehat{=}} C_1^j {\mathcal T}_j^j + 
C_1^p {\mathcal T}_p^j  \nonumber \\
{\mathcal T}_1^p & = & - \frac{\delta \Sigma}{\delta \left( \nabla_p X^1\right)} \mathrel{\widehat{=}} C_1^j {\mathcal T}_j^p + C_1^p {\mathcal T}_p^p.
\end{eqnarray}
These components obey the conservation equations $\nabla_j {\mathcal T}_0^j  + \nabla_p {\mathcal T}_0^p = 0$ and
$\nabla_j {\mathcal T}_1^j  + \nabla_p {\mathcal T}_1^p = 0$. 

\section{Continuous limit}

The continuous limit of $2D$ DTQWs is analyzed in detail in \cite{DBD14a}. The most interesting case is the one where the limit exists and coincides with the free Dirac equation, which is obtained by collecting all first order terms in the infinitesimal $\epsilon$ which fixes the time- and spatial-step. At first order in $\epsilon$, the operators $\nabla_j$ and $\nabla_p$ then tend towards $\epsilon \partial_t$ and $\epsilon \partial_x$, the operator $C$ tends towards unity and the operator $W$ tends towards $1 + i \epsilon m \sigma_1$ where $\sigma_1$ is 
the first Pauli matrix and $m$ plays the role of the fermion mass. Thus, at first order in $\epsilon$, $M^1_j$ coincides with the mass term of the Dirac equation while both $M^2_j$ and 
$M^3_j$ vanish at the same order. As for the kinetic terms, ${\bar K}_j$ coincides at first order in $\epsilon$ with the kinetic term of the Dirac equation while all other kinetic contributions vanish at that order.

\section{Conclusion}

We have shown that the quantum automata linear unitary dynamics derives from a Hamiltonian preserving basic action principle. 
For DTQWs with constant coefficients on the line, this basic principle can be extended by adding space-time coordinates as new variables defined on the space-time grid and the equations of motion for these new variables deliver energy and momentum conservation in the grid reference frame in the form of finite differences conservation equations obeyed by the energy and momentum densities together with their flux densities. We have finally proposed a manifestly covariant form of this extended action principle and computed the stress-energy of the DTQW in an arbitrary reference frame. The Lorentz invariance of the $2$D Dirac equation has been revisited in Appendix 3, while Appendix 2 shows how charge conservation for DTQWs can be recovered from the action principle. Appendix 1 presents the basics of discrete action principles on simple examples from classical mechanics. Though Appendix 1 is presented as a tutorial, all the material offered in it is new.

Let us now comment on these results and mention a few extensions. The first, very general conclusion of this manuscript is that the dynamics of several physically interesting discrete dynamical systems like quantum automata and in particular DTQWs does derive from action principles. The perhaps unexpected point is that, in the discrete case, the existence of an autonomous {\sl i.e.} time-independent Lagrangian does not guarantee energy conservation, except in very particular systems like quantum automata. For DTQWs, the Hamiltonian obtained from the basic action principle introduced in this article is actually the imaginary exponential on the so-called pseudo Hamiltonian already introduced in the literature independently of action principles. The momentum of the DTQW is essentially (twice) the sine of the wave-number. These two quantities represent physically the energy and the momentum of the DTQW in what we have called the grid reference frame. Associated spatial densities and currents can be also computed by choosing the right variables in the action.

The basic action principle valid for general quantum automata, as specialized to DTQWs, presents two limitations. First, the action involves only the grid coordinates $(j, p)$ and does not involve arbitrary space-time coordinates, so Lorentz invariance cannot even be discussed properly. Second, energy conservation is not generically valid for discrete action principles. Thus, extending the DTQW action to describe for example the dynamics of gauge fields acting on DTQWs would lead to a global dynamics whose Lorentz covariance cannot be discussed and which does not conserve any energy.

As explained in Appendix 1, the key to solving the second problem lies in introducing space-time coordinates $X^0$ and $X^1$ as new variables of a first extended action. The equations of motion then have to be solved, not only for the spinor $\Phi$ (and possible other fields like gauge fields if needed), but also for the space-time coordinates $X^0$ and $X^1$. Conservation equations for an energy and a momentum for the spinor (and possible other fields) then appear as the equations of motion for $X^0$ and $X^1$. Thus, in a generic case, space-time coordinates cannot be chosen freely and the solution obeying the equations of motion ensures that an energy and momentum for the remaining variables (spinor etc.) is conserved. In the simple case envisaged in this article {\sl i.e.} free DTQWs, the original dynamics does conserve energy and momentum. Thus, the equations of motion for the spinor and the space-time coordinates are solved identically by the original DTQW dynamics and the choice $X^0 = j$ and $X^1 = p$. As it stands, this first extension of the basic action principle does not bring a lot on table when only DTQWs are considered. But this first extended action can be generalized to describe inhomogeneous DTQWs and a coordinate-dependent kinetic term for other fields like gauge fields can also be added. The resulting equations of motion for the fields combined with the space-time coordinates will then enforce energy and momentum conservation. Thus, the first extension of the basic action solves the second problem above.

Note that, at this stage, the coordinate $X^0$ and $X^1$ are only arbitrary off-shell, and the coordinate-independent on-shell action obtained by replacing $X^0$ by $j$ and $X^1$ by $p$ is identical to the basic action obeyed by DTQWs. Thus, at this stage, the coordinates $(X^0, X^1)$ are not coordinates in an arbitrary Lorentz frame, but space-time coordinates in the Lorentz frame of the grid. They are arbitrary variables of the extended action and become equal to the grid coordinates by virtue of the equations of motion. Thus, this first extension of the basic action principle still describes physics in the grid reference frame only.

Change of reference frame can be performed by considering a second extension of the basic action, which is obtained by rewriting the first extension into a manifestly covariant manner. Let us now discuss some aspects of the quite involved computation. 

Invariance properties of the Dirac equation are notoriously non trivial conceptually. The Lorentz invariance of the $2$D flat space-time Dirac equations is reviewed in Appendix 3, with special emphasis on geometrical aspects. One of the non-standard aspects of the presentation is that flat-space time is viewed as a special case of the more general curved space-time. 

Under a change of Lorentz frame, wave functions of DTQWs and usual Dirac spinors transform the same way. The same goes for spin-operators acting on DTQWs and Dirac spinors, which all obey what we call the SOLT law. Both actions contain the same basic kinetic term, but the discrete action contains extra kinetic and mass terms which vanish at the continuous limit. The extra kinetic and mass terms can be written in a manifestly covariant manner which involves the time-like $2$-velocity of the grid in space-time and its orthogonal vector field or, if one prefers, {\sl i.e.} the projector unto the space orthogonal to the velocity field. This sort of dependence is standard when one describes in a covariant manner situations which exhibit a special reference frame. Standard examples are relativistic fluid dynamics, including acoustics and diffusive transport in a relativistic fluid, cosmological structure formation and relativistic Hamiltonian formulations. Indeed, scalar equations of state for relativistic fluids 
\cite{dGvLvW, Israel87a} link energy density and pressure in the local rest-frame of the fluid and this quantities are contractions of the stress-energy tensor of the fluid with its local velocity field and with the projector orthogonal to this field. Similarly, relativistic acoustics \cite{LandauHydro, DB97a} studies perturbations around a given flow characterized notably by its velocity field. Also, the stochastic equations defining physically realistic relativistic stochastic processes \cite{CD08a} can be written in a manifestly covariant manner which involves the velocity of the fluid in which diffusion takes place. In the same vein, cosmological structure formation \cite{Peeb93a} relies on a perturbation expansion around an isotropic homogeneous solution of Einstein equation, which exhibits a preferred reference frame \cite{HE73a}. This frame is also identified with the so-called co-moving frame where the CMB appears isotropic \cite{KT94a}. Finally, the Hamiltonian formulation of any relativistic theory, from point-particle dynamics \cite{Barut80a} to General Relativity \cite{Wald84a} is naturally perfectly covariant and is based on a foliation of space-time, which amounts to choosing a reference frame at each point {\sl i.e.} a velocity field. 


Geometrically speaking, the picture that emerges from this article can be summed up in the following manner. For the Dirac equation, the base space-time is a differential manifold, which is a continuous set of points and coordinate systems are used to label these points by real numbers. For the DTQWs, the base space-time is the grid and coordinates are used to label the discrete points of this grid by real numbers. Both dynamics derive from an action principles which can be 
written in a manifestly covariant manner. The Dirac action principle does not involve a particular velocity field $U$ {\sl i.e.} it does not involve any preferred reference frame. On the contrary, the DTQW action principle, when written in arbitrary Lorentz coordinates, does involve a quantity which can be interpreted as the velocity of the grid in space-time and whose components become $(1, 0)$ if one uses coordinates attached to the grid.

Let us conclude by mentioning a few natural extensions of this work. The analysis presented in this article should be carried out on more general DTQWs with constant coefficients, not only walks in higher dimensions, but also several step walks. One should then address walks with non constant coefficients, which are essentially discrete models of spinors coupled to gauge fields, and then proceed by adding to the 
action Lorentz invariant kinetic terms for the gauge fields, thus obtaining energy and momentum conserving discrete models of spinors self-consistently coupled to gauge fields. Finally, action principles can also be used in path-integrals to second quantize DTQWs.

\section{\label{sec:app1}Appendix 1: Toolbox for discrete action principles}

\subsection{Basics}

For simplicity, we consider an action $S$ which is a functional of a single real variable $q$ defined on the discrete `time' line $\mathbb{Z}$. This action can be viewed as as function of the variables $q = (q_j)_{j \in \mathbb{Z}}$. The corresponding equations of motion are thus obtained by enforcing that $S$ is an extremum for arbitrary independent variations of the $q_j$'s. The equations of motion thus read $\forall j \in \mathbb{Z}, \partial S/\partial q_j = 0$.

Suppose now that the explicit expression of $S$ in terms of the $q_j$ makes it natural to write $S$ as a function $\tilde S$ of both the $q_j$'s and the finite differences $v_j = q_{j + 1} - q_j = (Dq)_j$, as is for example the case with the `mechanical' action
\begin{eqnarray}
S_m (q) & = & \sum_j [ \frac{1}{2} \left(q_{j + 1} - q_j \right)^2 - \phi(q_j)] \nonumber \\
& = & \sum_j [ \frac{1}{2} v_j^2 - \phi(q_j)] \nonumber \\
& = & {\tilde S}_m (q, v).
\end{eqnarray} 
The equations of motion can easily be recovered from $\tilde S$. Indeed, 
\begin{eqnarray}
\frac{\partial S}{\partial q_j} & = & \frac{\partial {\tilde S}}{\partial v_j} \frac{\partial v_j}{\partial q_j} +  \frac{\partial {\tilde S}}{\partial v_{j-1}} \frac{\partial v_{j-1}}{\partial q_j} + \frac{\partial {\tilde S}}{\partial q_j} \nonumber \\
& = & - \frac{\partial {\tilde S}}{\partial v_j} + \frac{\partial {\tilde S}}{\partial v_{j-1}} + \frac{\partial {\tilde S}}{\partial q_j}.
\end{eqnarray}
Introducing the momentum 
\begin{equation}
p_j = \frac{\partial {\tilde S}}{\partial v_j},
\end{equation}
the equation of motion $\partial S/\partial q_j = 0$ thus reads
\begin{equation}
(Dp)_{j - 1} = p_j - p_{j - 1}= \frac{\partial {\tilde S}}{\partial q_j},
\label{eq:motion1}
\end{equation}
which closely parallels the usual Euler-Lagrange equation.

Suppose now that one chooses for example $v'_j = q_{j+1} - q_{j-1} = (D' q)_j$ as velocity and work with an ${\tilde S}' (v', q)$. A similar computation delivers the equation of motion for $q_j$ under the form
\begin{equation}
(D'p')_{j} = p'_{j+1} - p'_{j - 1}= \frac{\partial {\tilde S}'}{\partial q_j}
\end{equation}
with 
\begin{equation}
p'_j = \frac{\partial {\tilde S}'}{\partial v'_j}.
\end{equation}

If one of the $q_j$'s is a cyclic variables of ${\tilde S}$ ({\sl resp.} ${\tilde S}'$), its conjugate momentum $p_j$ ({\sl resp.} $p'_j$) is 
conserved in the sense that $(Dp)_j = 0$ ({\sl resp.} $(D'p')_j = 0$). 

\subsection{Hamiltonian}

Consider now the Legendre transform ${\tilde S}_H = \sum_j p_j  v_j - {\tilde S}$ of $\tilde S$. By definition, ${\tilde S}_H$ is a function of the $(q_j, p_j)$'s and its 
differential reads
\begin{equation}
d{\tilde S}_H = \sum_j [ - \frac{\partial {\tilde S}}{\partial q_j} dq_j + v_j dp_j ].
\end{equation}
Combining this with the definition of $v_j = q_{j+1} - q_j$ and with the equation of motion (\ref{eq:motion1}) leads to
\begin{eqnarray}
\label{eq:Ham1}
(Dp)_{j-1} & = & - \frac{\partial {\tilde S}_H}{\partial q_j}\nonumber \\
(Dq)_j & = & \frac{\partial {\tilde S}_H}{\partial p_j}.
\end{eqnarray}
One can show in a similar way that 
\begin{eqnarray} 
\label{eq:Ham2}
(D'p')_{j} & = & - \frac{\partial {\tilde S}'_H}{\partial q_j}\nonumber \\
(D'q)_j & = & \frac{\partial {\tilde S}'_H}{\partial p_j}
\end{eqnarray}
where  ${\tilde S}'_H$ is the Legendre transform of  ${\tilde S}'$ with respect to the $v'_j$'s.

Suppose now that the action $S$ does not depend explicitly on $j$ and that ${\tilde S}_H (p, q) = \sum_j H (p_j, q_j)$. Even then, 
the discrete Hamiltonian (or symplectic) equations do {\sl not} generally imply energy conservation. Take for example the action $S_m$, 
for which ${\tilde S}_{mH} (q, p) = \sum_j [p_j^2/2 + \phi (q_j)]$ which generates for all $j$ the equations of motion $(Dp)_{j-1} = - \partial \phi /\partial q_j$ and 
$(Dq)_j = p_j$. Then, 
\begin{equation}
H_{j+1} - H_j = \phi(q_j + (Dq)_j) - \phi(q_j) - p_j \left( \frac{\partial \phi}{\partial q} \right)_{j + 1} + \frac{1}{2} \  \left( \frac{\partial \phi}{\partial q} \right)_{j + 1}^2,
\end{equation}
which only vanishes for constant potentials $\phi$ {\sl i.e.} for free particles. Thus, $S_m$ is a symplectic discretization of the motion of a point particle in the potential $\phi$, but this discretization only conserves energy for free particles.

\subsection{Conserved energy}

For a large class of actions, the above non conservation of energy can be remedied by introducing a new `time' variable $t_j$.

Consider first for example an autonomous action of the form 
${\tilde S} (q, v) = \sum_j {\tilde s} (q_j,v_j)$ and build from $\tilde S$ the following action 
\begin{equation}
{\tilde \Sigma} (q, t, v, V) = \sum_j {\tilde \sigma}(q_j, v_j/V_j) V_j = \sum_j {\tilde \sigma}(q_j, u_j) V_j,
\end{equation}
where $V_j = (Dt)_j$. In passing from ${\tilde S}$ to ${\tilde \Sigma}$, $\Delta j = (j + 1) - j = 1$ has been formally replaced by $\Delta t = t_{j + 1} - t_j =  V_j$ and the velocity $v_j = Dq_j/(\Delta j)$ has been replaced by $u_j = Dq_j/(\Delta t) = v_j/V_j$.

The momenta $\pi_j$ and $\Pi_j$ conjugate to $q_j$ and $t_j$ read:
\begin{eqnarray}
\label{eq:motion3}
\pi_j & = & \frac{\partial {\tilde \Sigma}}{\partial v_j} = V_j \left( \frac{\partial {\tilde \sigma}}{\partial u} \right)_{u_j = v_j/V_j} \nonumber \\
\Pi_j & = & \frac{\partial {\tilde \Sigma}}{\partial V_j} = {\tilde \sigma}(u_j, q_j) - u_j \  \left( \frac{\partial {\tilde \sigma}}{\partial u} \right)_{u_j = v_j/V_j}
\end{eqnarray}
The Legendre transform of ${\tilde \Sigma}$ with respect to the $q$'s and $t$'s vanishes identically {\sl i.e.} there is no Hamiltonian from which to derive the equations 
of motion. On the other hand, the time $t$ is a cyclic variable of the action, so its momentum $\Pi$ is conserved ${\sl i.e.}$ $\Pi_j$ does not depend on $j$. Now, the second equation in (\ref{eq:motion3}) shows that $\Pi_j$ is the Legendre transform of ${\tilde S}$ with respect to the velocity $u_j$ and, thus, represents the energy at time $t_j$. One thus gets energy conservation as the equation of motion for the time variable $t$ and the actual equation of motion for $q$ is
\begin{equation}
(D p)_{j - 1} = V_j \left( \frac{\partial {\tilde \sigma}}{\partial q} \right)_{q_j}.
\end{equation}

Taking again the action $S_m$ as an example,
\begin{equation}
{\tilde \Sigma}_m (q, t, v, V) = \sum_j [ \frac{1}{2} \left(\frac{v_j}{V_j}\right)^2 - \phi(q_j)] V_j,
\end{equation}
\begin{eqnarray}
\label{eq:motion3m}
\pi_j & = & \frac{v_j}{V_j} = \frac{(Dq)_j}{(Dt)_j} \nonumber \\
\Pi_j & = & -\frac{1}{2} \left( \frac{(Dq)_j}{(Dt)_j} \right)^2 - \phi (q_j)
\end{eqnarray}
and the equation of motion for $q_j$ reads
\begin{equation}
\left(D \, \frac{Dq}{Dt}\right)_{j - 1} = - (D t)_j \left( \frac{\partial \phi}{\partial q} \right)_j.
\end{equation}

Suppose now that the function $\tilde \sigma$ and, thus, the action ${\tilde \Sigma}$ depend explicitly on $t$:
\begin{equation}
{\tilde \Sigma} (q, t, v, V) = \sum_j {\tilde \sigma}(t_j, q_j, v_j/V_j) V_j = \sum_j {\tilde \sigma}(t_j, q_j, u_j) V_j,
\end{equation}
The equation of motion for $q$ is not modified but the equation of motion for $t$ now reads:
\begin{equation}
(D \Pi)_{j - 1} = \left(\frac{\partial {\tilde \sigma}}{\partial t}\right)_{j}.
\end{equation}
The energy is not conserved because the action itself depends on $t$. This kind of action is useful in describing a system (variable $q$) evolving under the influence of an external ({\sl i.e.} imposed) non constant force field. Note that a complete consistent treatment including both $q$ and the field as dynamical variables only involves 
time (and space-) independent Lagrangian (densities). Note also that, in general, the natural time variable for $\tilde \sigma$ is $t$, not the iteration index $j$.

\section{\label{sec:app2}Appendix 2: Charge Conservation for DTQWs on the line}
Let us first consider, for simplicity sakes, the so-called one-step $(1 + 1)$D DTQWs with two component wave functions $\Psi = (\psi^-, \psi^+)^\dagger$, for which $U_j = V_j T$ where $T$ is the $j$-independent spatial translation operator $T$ defined by
\begin{eqnarray}
(T\psi)^{-}_{p} & = &  \psi^{-}_{p+1} \nonumber \\
(T \psi)^{+}_{p} & = &  \psi^{+}_{p-1} 
\end{eqnarray}
and $V_j$ is defined 
\begin{equation}
(V_j \Psi)_p = W_{j, p} \Psi_p
\end{equation}
with $W_{j,p}$ an arbitrary $j$- and $p$-dependent operator in $U(2)$. 

Let $(\rho^\pm, \delta^\pm)$ be the modulus and phase of the component $\psi^\pm$ and introduce the new phases $\mu= (\delta^+ + \delta^-)/2$ and $\delta = (\delta^+  -\delta^-)/2$. Since $\phi^\pm = (\mu \pm \delta)$, the two-component spinor thus reads
\begin{equation}
\left( \begin{array}{c} \psi^{-}\\ \psi^{+} \end{array} \right)= e^{i \mu} \left( \begin{array}{c} \rho^{-} e^{- i \delta } \\ \rho^{+} e^{+  i \delta}\end{array} \right)
\end{equation}
The equation of motion for each of the $\mu_{j, p}$'s reads $\partial S/\partial \mu_{j, p} = 0$.

We will now show that the action (\ref{eq:Mainaction}) does not involve the variables $\mu_{j, p}$ $\sl{per se}$ but only linear combinations which are discrete equivalents of their time- and space-derivatives. The contribution $\sum_j < \Psi_{j +1} \mid \Psi_{j +1}>$ is trivially independent of the $\mu_{j, p}$'s so let us focus on $\sum_j < \Psi_{j +1} \mid U_j \Psi_{j}>$.
Now,
\begin{eqnarray}
(T \Psi_j)_{j, p} & = & 
\left( \begin{array}{c} \psi^{-}_{j, p+1}\\ \psi^{+}_{j, p-1} \end{array} \right) \nonumber \\
& = & e^{i {\tilde \mu}_{j, p}} 
\left( \begin{array}{c} \rho^{-}_{j, p+1} e^{- i {\tilde \delta}_{j, p}}\\ \rho^{+}_{j, p-1} e^{+ i {\tilde \delta}_{j, p}}\end{array} \right)
\label{eq:TPsi}
\end{eqnarray}
where 
\begin{equation}
{\tilde \mu}_{j, p} = \frac{\mu_{j, p+1} + \mu_{j, p-1} - \delta_{j, p+1} + \delta_{j, p-1}}{2}
\end{equation}
and
\begin{equation}
{\tilde \delta}_{j, p} = \frac{- \mu_{j, p+1} + \mu_{j, p-1} + \delta_{j, p+1} + \delta_{j, p-1}}{2}.
\end{equation}
It follows that the action (\ref{eq:Mainaction}) only depends on the $\delta_{j, p}$'s and on 
\begin{equation}
\left(\nabla_p \mu \right)_{j, p} = \frac{\mu_{j, p+1} - \mu_{j, p-1}}{2}
\end{equation}
and
\begin{equation}
\left(\nabla_j \mu \right)_{j, p} = \mu_{j+1, p} - \frac{\mu_{j, p+1} +\mu_{j, p-1}}{2}.
\end{equation}

The variables $\mu_{j, p}$ are thus cyclic. There is therefore a corresponding conserved charge and the associated conserved current $(J_j, J_p)$ can be computed from the action (\ref{eq:Mainaction}) by the same method as the one described in the Appendix on the simpler case where all quantities depend only on the discrete time $j$. Indeed, the above equation of motion for $\mu_{j, p}$ can be written as
%
\begin{equation}
\left(\nabla'_j J_j \right)_{j, p} + \left(\nabla_p J_p \right)_{j, p} = 0
\end{equation}
where
\begin{equation}
\left(\nabla_p J_p \right)_{j, p} = \left(J_p \right)_{j, p + 1} -  \left(J_p \right)_{j, p - 1},
\end{equation}
\begin{equation}
\left(\nabla'_j J_j \right)_{j, p} = \frac{\left(J_j \right)_{j, p + 1} +  \left(J_j \right)_{j, p - 1}}{2} - \left(J_j \right)_{j -1, p}
\end{equation}
and
\begin{equation}
\left(J_p\right)_{j, p} = \frac{\partial S}{\partial \left(\nabla_p \mu \right)_{j, p}},
\end{equation}
\begin{equation}
\left(J_j\right)_{j, p} = \frac{\partial S}{\partial \left(\nabla_j \mu\right)_{j, p}}.
\end{equation}
In the last two equations, $S$ stands for the action (\ref{eq:Mainaction}) understood as a function of the
$\left(\nabla_j \mu \right)_{j, p}$'s and $\left(\nabla_p \mu\right)_{j, p}$'s, and not of the $\mu_{j, p}$'s. 

A direct computation leads to 
\begin{eqnarray}
(J_j)_{j, p} & = & +  (\rho^-_{j, p})^2 + (\rho^+_{j, p})^2 \nonumber \\
(J_p)_{j, p} & = & -  (\rho^-_{j, p})^2 + (\rho^+_{j, p})^2,
\end{eqnarray}
which are the usual expressions for the spatial density and current of the DTQW \cite{AF16b}.

%
%
%

%
%
%


\section{\label{sec:app3}Appendix 3: Lorentz invariance of the $2D$ flat space-time Dirac action}

In an arbitrary given Lorentz frame $\mathcal F$, the flat space-time $2D$ Dirac Lagrangian density reads \cite{SR94a}
\begin{equation}
L \left[ \Psi, \Psi^\dagger \right] = \eta_{\sigma \delta} (\Psi^\delta)^* \left( e^\mu_a \gamma^0 \gamma^a \partial_\mu + i m \gamma^0 \right)^\sigma_\omega \Psi^\omega 
\label{eq:DiracA1}
\end{equation}
where the $\Psi^\sigma$'s 
are the component of $\Psi$ 
on the spin space basis $(b_-, b_+)$ of $\mathcal F$ and the components of the metric $\eta$ in spin space are $\eta_{\sigma \delta} = 1$ if $\sigma = \delta$ and $0$ otherwise. The $e^\mu_a$'s are the $2$-bein coefficients which we choose to be $e^t_0 = e^x_1 = 1$, $e^t_1 = e^x_0 = 0$ and we 
retain a representation where 
\begin{eqnarray}
(\gamma^0)^\sigma_\delta & = & 
\begin{pmatrix} 
0 & 1 \\
1 & 0 
\end{pmatrix}
\nonumber \\
(\gamma^1)^\sigma_\delta & = &
\begin{pmatrix} 
0 & 1 \\
-1 & 0 
\end{pmatrix},
\end{eqnarray}
so that $(\gamma^0)^2 = - (\gamma^1)^2 =1$, $\gamma^0 \gamma^1 = \mbox{diag} (- 1, 1) = - \sigma_3$ and $(\eta. \gamma_0)_{\delta \omega} = \eta_{\sigma \delta} (\gamma^0)^\sigma_\omega = 1$ if $\delta \ne \omega$ and $0$ otherwise. Making all this explicit in the Dirac Lagrangian delivers:
\begin{eqnarray}
L \left[ \Psi, \Psi^\dagger \right] & = & (\Psi^{-})^* \left( + \partial_t - \partial_x \right ) \Psi^{-}   \nonumber \\
&  & +    (\Psi^{+})^* \left(+ \partial_{t} + \partial_{x} \right ) \Psi^{+}  \nonumber \\
& &  +  i m \left[ (\Psi^{-})^*  \Psi^{+} + (\Psi^{+})^*  \Psi^{-} \right].
\label{eq:Lag2}
\end{eqnarray}

Suppose now one performs a change of basis in spin-space and define the new basis vectors 
$b_{\sigma'}$ by $b_{-'} = \lambda b_{-}$, $b_{+'} = b_{+}/\lambda$, where $\lambda$ is a non-vanishing real constant. The components of $\Psi$ in this new basis are $\Psi^{-'} = \Psi^{-}/\lambda$ and $\Psi^{+'} = \lambda \Psi^{+}$ and the operators $\gamma^0$ and $\gamma^1$ are represented by the matrices
\begin{eqnarray}
(\gamma^0)^{\sigma'}_{\delta'} & = & 
\begin{pmatrix} 
0 & \lambda^{-2} \\
\lambda^2 & 0 
\end{pmatrix}
\nonumber \\
(\gamma^1)^{\sigma'}_{\delta'} & = &
\begin{pmatrix} 
0 & \lambda^{-2} \\
- \lambda^2 & 0 
\end{pmatrix}.
\end{eqnarray}
Similarly, $\eta_{-'-'} = \lambda^2$, $\eta_{+'+'} = \lambda^{-2}$ and the other components vanish.
This leaves the matrix representations of $(\gamma^0)^2 = 1$, $(\gamma^1)^2 = - 1$, $\gamma^0 \gamma^1$ and $\eta. \gamma^0$ unchanged. 

The basis change thus preserves the Clifford algebra and therefore amounts to a change of representation. One obtains, either from (\ref{eq:Lag2}) or 
directly from (\ref{eq:DiracA1}) leaving the $2$-bein unchanged:
\begin{eqnarray}
L \left[ \Psi, \Psi^\dagger \right] & = & \lambda^2 (\Psi^{-'})^* \left( + \partial_t - \partial_x \right ) \Psi^{-'}   \nonumber \\
&  & +  \frac{1}{\lambda^2}  (\Psi^{+'})^* \left(+ \partial_{t} + \partial_{x} \right ) \Psi^{+'}  \nonumber \\
& &  +  i m \left[ (\Psi^{-'})^*  \Psi^{+'} + (\Psi^{+'})^*  \Psi^{-'} \right].
\label{eq:Lagprime1}
\end{eqnarray}
This is identical to (\ref{eq:Lag2}), except for the $\lambda^{\pm 2}$ factors. These factors come from the components of the spin metric $\eta$ in the new basis (remember that the matrix representing $(\gamma^0)^2 = 1$ and $\gamma^0 \gamma^1$ are unchanged by the change of basis). But, looking only at 
(\ref{eq:Lagprime1}) and (\ref{eq:DiracA1}) together, one could be tempted to interpret these two $\lambda^{\pm 2}$ factors as coming from the $2$-bein, and not the components of the metric $\eta$.



%
%

Indeed, consider the new spin-metric $\eta'$ defined by its components $\eta'_{\sigma' \delta'} = 1$ if $\sigma' = \delta'$ and $0$ otherwise, the new operators $(\gamma')^0$ and $(\gamma')^1$ defined by 
\begin{eqnarray}
((\gamma')^0)^{\sigma'}_{\delta'} & = & 
\begin{pmatrix} 
0 & 1 \\
1 & 0 
\end{pmatrix}
\nonumber \\
((\gamma')^1)^{\sigma'}_{\delta'} & = &
\begin{pmatrix} 
0 & 1 \\
- 1 & 0 
\end{pmatrix},
\end{eqnarray}
and the new $2$-bein defined by its components on the coordinate basis $(\partial_t, \partial_x)$:
\begin{eqnarray}
(e')^t_0 & = &  (+ \lambda^2 + \lambda^{-2} )/2 \nonumber \\
(e')^x_0 & = &  (- \lambda^2 + \lambda^{-2} )/2 \nonumber \\
(e')^t_1 & = &  (- \lambda^2 + \lambda^{-2} )/2 \nonumber \\
(e')^x_1 & = &  (+ \lambda^2 + \lambda^{-2} )/2 \nonumber \\
\end{eqnarray}
The triplet $(\eta', \gamma', e')$ delivers the same Lagrangian (\ref{eq:Lagprime1}) as the triplet $(\eta, \gamma, e)$.

%


Now, we have not changed space-time coordinates so far and the new $2$-bein $e'$ has been introduced by its components on the original tangent basis $\partial_\mu$. As well known, it is possible to introduce new 
coordinates $x^{\mu'}$ 
and the associated coordinate basis $\partial_{x^{\mu'}}$ where the non vanishing of the components of the $2$-bein $e'$ 
are $(e')^{t'}_0 = (e')^{x'}_1 = 1$. Using these
coordinates, the Dirac Lagrangian reads
\begin{eqnarray}
L \left[ \Psi, \Psi^\dagger \right] & = &  (\Psi^{-'})^* \left[\left( + \partial_{t'} - \partial_{x'} \right ) \Psi^{-'}  + i m \Psi^{+'} \right]  \nonumber \\
&  + & (\Psi^{+'})^* \left[\left(+ \partial_{t'} + \partial_{x'} \right ) \Psi^{+'}  + i m \Psi^{-'} \right], 
\label{eq:Lagprime2}
\end{eqnarray}
which is formally identical to (\ref{eq:Lag2}).


Note that, at fixed mass $m$, the change of $\gamma$ operators performed above may be considered as collateral to, or induced by the change of spin metric. Indeed, at fixed value of $m$, keeping the mass term fixed comes down to keeping $\eta. \gamma$ constant and thus, changing $\eta$ makes it necessary to change the $\gamma$ operators.
Note also that the $2$-bein defines the space-time metric, so a change of $2$-bein is actually a change of space-time metric. 
A Lorentz transformation is thus basically a change of spin basis compensated by or compensating a change of coordinate basis/coordinate system through a combined change of spin metric (together with the induced change of $\gamma$ operators) and space-time metric.  

All this can be seen in a perhaps more compact way. Consider expressions of the form
\begin{equation}
F \left[ \Psi, \Psi^\dagger \right] = \eta_{\sigma \delta} (\Psi^\delta)^* O^\sigma_\omega \Psi^\omega.
\end{equation}
The Lagrangian (\ref{eq:DiracA1}) corresponds to $O_D = e^\mu_a \gamma^0 \gamma^a \partial_\mu + i m \gamma^0$. Consider the same two spin bases and the same two spin metrics as above. It is simple to check that $F$ can be rewritten as
\begin{equation}
F \left[ \Psi, \Psi^\dagger \right] = (\eta')_{\sigma' \delta'} (\Psi^{\delta'})^* (O')^{\sigma'}_{\omega'} \Psi^{\omega'}
\end{equation}
with
\begin{eqnarray}
O^{L'}_{L'} & = & \lambda^{+2} O^L_L \nonumber \\
O^{R'}_{R'} & = & \lambda^{-2} O^R_R \nonumber \\
O^{L'}_{R'} & = & O^L_R \nonumber \\
O^{R'}_{L'} & = & O^R_L.
\end{eqnarray}
We call this the Spin Operator Lorentz Transformation (SOLT) law. Now apply the SOLT law to the mass term in $O_D$. The operator $\gamma^0$ is represented by 
an anti-diagonal matrix so, at constant $m$, the components of $(\gamma')^0$ in the prime basis must be the same as those of $\gamma^0$ in the original basis. The Clifford algebra then fixes $(\gamma')^1$. One remarks that the prime components of $(\gamma')^a$ are identical to the original components of $\gamma^a$.

If one now turns to the kinetic term in $O_D$, both operators $(e^\mu_0 \partial_\mu) (\gamma^0 \gamma^0)$ and $(e^\mu_1 \partial_\mu) (\gamma^0 \gamma^1)$ are represented by diagonal matrices in the original basis. The SOLT law thus generates extra $\lambda^{±\pm 2}$ factors, which do not come from the $(\gamma')^a$. These extra factors must therefore come from the two factors (scalars in spin space) $(e^\mu_a \partial_\mu)$. This determines the change of $2$-bein and, in turn, the change of coordinates (to keep the coordinate components of the $2$-bein unchanged). 

The SOLT law is used extensively in Section ... of this article, where the Lorentz invariance of the action for DTQWs is discussed.

We finally recall that 
\begin{equation}
(e')^{\mu'}_a = \frac{\partial x^{\mu'}}{\partial x^\nu} (e')^{\nu}_a = N^{\mu'}_\nu (e')^{\nu}_a
\end{equation}
so that 
\begin{equation}
(e')^{\mu}_a = (N^{-1})^\mu_{\nu'} (e')^{\nu'}_a,
\label{eq:change2bein}
\end{equation}
where the components of $N^{-1}$ are given by
\begin{eqnarray}
(N^{-1})^t_{t'} & = & +  \frac{1}{\delta}  \frac{\partial x'}{\partial x} \nonumber \\
(N^{-1})^t_{x'} & = &  - \frac{1}{\delta} \frac{\partial t'}{\partial x} \nonumber \\  
(N^{-1})^x_{t'} & = &  - \frac{1}{\delta}  \frac{\partial x'}{\partial t}  \nonumber \\
(N^{-1})^x_{x'}& = &  + \frac{1}{\delta} \frac{\partial t'}{\partial t}
\end{eqnarray}
with
\begin{equation}
\delta = \frac{\partial t'}{\partial t} \frac{\partial x'}{\partial x} - \frac{\partial t'}{\partial x} \frac{\partial x'}{\partial t}.
\end{equation}

At fixed values of the components $(e')^{\nu'}_a$, the values of the components $(e')^{\mu}_a$ can thus be encoded in the functions $t'(t, x)$ and $x'(t, x)$. This is used in the main part of this article to introduce coordinates in the DTQW action.



\end{document}